\documentclass{optica-article}

\journal{opticajournal} % for journals or Optica Open

\articletype{Research Article}
\usepackage{physics}
\usepackage{lineno}
\begin{document}

\title{Linear-optical generation of approximate higher-component cat states}

\author{Shohei Kiryu,\authormark{1,*} and Masahiro Takeoka\authormark{1,2}}

\address{\authormark{1}Department of Electronics and Electrical Engineering, Keio University, 3-14-1 Hiyoshi, Kohoku-ku, Yokohama 223-8522, Japan\\
\authormark{2}National Institute of Information and Communications Technology (NICT), Koganei, Tokyo 184-8795, Japan\\
}

\email{\authormark{*}kiryu.opt@keio.jp} %% email address is required; see note below about the corresponding author designation

\begin{abstract*}
    We propose a method for generating higher-component cat states by interfering two $N$-component cat states on a half beam splitter and measuring one output with a photon-number-resolving detector. The Fock components of the conditional output are restricted to a single residue class modulo $2N$. For $N=3$, an ideal six-component cat state can be generated exactly, while for arbitrary $N$, the conditional output approaches an ideal $2N$-component cat state for sufficiently large input amplitudes.
\end{abstract*}

\section{Introduction}
    Photonic quantum states are expected to have a wide range of applications in quantum information processing, including quantum computation, quantum communication, and quantum sensing~\cite{takeda2019toward}. Quantum error correction is required to protect quantum states from noise such as photon loss and dephasing. Bosonic codes enable hardware-efficient quantum error correction by encoding quantum information in a single bosonic mode~\cite{Gottesman2001GKP,ralph2003quantum,michael2016new}.

    Among bosonic codes, the Gottesman–Kitaev–Preskill (GKP) code is a promising approach to fault-tolerant photonic quantum computation~\cite{Gottesman2001GKP}. Ideal GKP states can correct small displacement errors, and Clifford operations can also be implemented relatively easily. On the other hand, their generation requires infinite squeezing, and only approximate states with finite squeezing can be generated experimentally. Furthermore, generating GKP states with sufficient quality for quantum information processing remains challenging~\cite{Konno2024GKP,larsen2025integrated,yu2026extensible}. 

    In contrast, cat codes are another class of bosonic codes based on superpositions of finite-energy coherent states and have been proposed for applications in quantum computation and quantum communication~\cite{ralph2003quantum,browne2005resource,grimsmo2020quantum,marinoff2024explicit,li2024performance,endo2025quantum,udupa2025performance}. 
    In optical systems, two-component cat states have been experimentally generated~\cite{gerrits2010generation,endo2023non,endo2026four}, and theoretical methods for generating four-component cat states and performing error correction using these states have also been proposed~\cite{hastrup2020deterministic,hastrup2022all,su2022universal,boon2026generalised}. 
    More recently, a Bell measurement for arbitrary $N$-component cat qubits using a half beam splitter (HBS) and photon-number-resolving detectors (PNRD) has been proposed, with potential applications in quantum information processing~\cite{oe2026linear}.

\begin{figure*}[t]
    \centering
    \includegraphics[width=\linewidth]{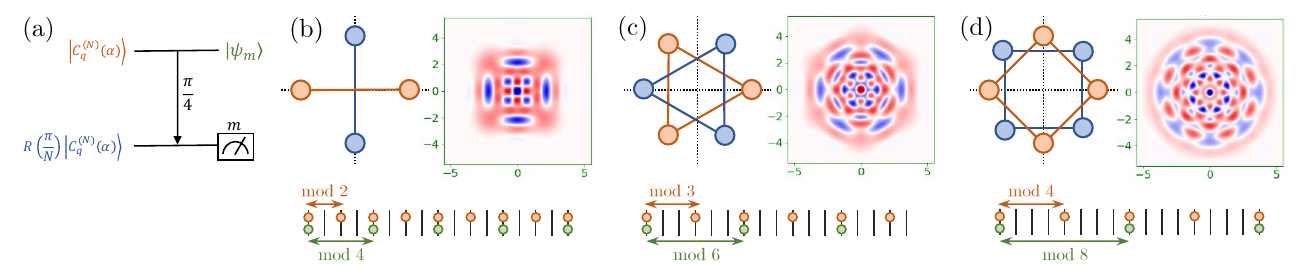}
    \caption{Generation scheme for higher-component cat states. (a) Generation circuit using an HBS and PNRD. Two $N$-component cat states are used as inputs, with one rotated by $\pi/N$. One output mode is measured by a PNRD, while the other is retained. (b)--(d) Illustrations for $N=2,3,4$. The blue and orange points represent the coherent-state components of the two inputs. The lower diagrams show the change in the allowed Fock components from modulo $N$ to modulo $2N$.}
    \label{FIG01_GraphicalAbstract}
    \end{figure*}

    In this paper, we show that a state with the same Fock support as a $2N$-component cat state can be generated by interfering two $N$-component cat states on an HBS and measuring one output with a PNRD. For $N=3$, our method can generate an ideal six-component cat state exactly. For arbitrary $N$, it can also approximately generate an ideal $2N$-component cat state when the input amplitude is sufficiently large. Fig.~\ref{FIG01_GraphicalAbstract} shows an overview of this method and the changes in the Fock components.

\section{Generation scheme}
    First, we define an $N$-component cat state. An $N$-component cat state with amplitude $\alpha$ can be expressed in the Fock basis as
    \begin{equation}
        \ket{C_q^{(N)}(\alpha)}\propto \sum_{\ell=0}^{\infty}\frac{\alpha^{N\ell+q}}{\sqrt{(N\ell+q)!}}\ket{N\ell+q}.
        \label{eq:N_component_cat_A_Fock_expansion}
    \end{equation}
    As shown in \eqref{eq:N_component_cat_A_Fock_expansion}, an $N$-component cat state consists only of Fock components with photon numbers equal to $q$ modulo $N$. In ~\eqref{eq:N_component_cat_A_Fock_expansion}, the case $q=0$ corresponds to a superposition of $N$ coherent states with equal coefficients. For $q\neq0$, the relative phases between the coherent-state components are different.

    Next, we define the input states. We consider two input modes, $A$ and $B$, and define the rotation operator as $R(\theta)=\exp(i\theta\hat{n})$. Here, $\hat{n}$ is the photon-number operator, and $\theta$ is the rotation angle. Using \eqref{eq:N_component_cat_A_Fock_expansion}, we define the initial states as
    \begin{subequations}\label{eq:N_component_cat_inputs}
        \begin{equation}
            \ket{\psi_A}=\ket{C_q^{(N)}(\alpha)}_A,
            \label{eq:N_component_cat_input_A}
        \end{equation}
        \begin{equation}
            \ket{\psi_B}=R_B(\pi/N)\ket{C_q^{(N)}(\alpha)}_B.
            \label{eq:N_component_cat_input_B}
        \end{equation}
    \end{subequations}
    For the input state in mode $B$ defined in \eqref{eq:N_component_cat_input_B}, the rotation operator acts on each Fock component as
    \begin{equation}
        R(\pi/N)\ket{N\ell+q}=e^{i\pi q/N}(-1)^{\ell}\ket{N\ell+q}.
        \label{eq:phase_rotation_N_component_Fock}
    \end{equation}
    In \eqref{eq:phase_rotation_N_component_Fock}, the factor $e^{i\pi q/N}$ is independent of $\ell$ and can be ignored as a global phase. On the other hand, the factor $(-1)^{\ell}$ changes the signs of the Fock coefficients depending on the parity of $\ell$. This phase structure contributes to the selection rule for the conditional output through interference at the HBS.

    We denote the two output modes of the HBS by $C$ and $D$. Mode $C$ is retained, and mode $D$ is measured. For coherent-state inputs with arbitrary amplitudes $\gamma$ and $\delta$, the output of the HBS is given by
    \begin{equation}
        U\ket{\gamma}_A\ket{\delta}_B=\ket{(\gamma+\delta)/\sqrt{2}}_C\ket{(\gamma-\delta)/\sqrt{2}}_D.
        \label{eq:HBS_coherent_state_transformation}
    \end{equation}

    Finally, we consider the projection corresponding to the PNRD measurement. We denote the detected photon number by $m$, a nonnegative integer. The conditional output state corresponding to the measurement outcome $m$ is given by
    \begin{equation}
        \ket{\widetilde{\psi}_m}\propto{}_D\bra{m}U\ket{\psi_A}\ket{\psi_B}.
        \label{eq:conditional_output_unnormalized}
    \end{equation}
    
    We analyze the conditional output state in \eqref{eq:conditional_output_unnormalized} in the Fock basis to determine which Fock components remain for the PNRD measurement outcome $m$. We first consider the constraint imposed by photon-number conservation and then the additional constraint due to interference. We denote the indices of the two input Fock components by $\ell_1$ and $\ell_2$ and define $L=\ell_1+\ell_2$. We denote the photon number in the retained mode $C$ by $p$. From photon-number conservation, we obtain
    \begin{equation}
        p=(N\ell_1+q)+(N\ell_2+q)-m=NL+2q-m.
        \label{eq:remaining_photon_number_after_detection}
    \end{equation}
    The allowed photon numbers $p$ therefore satisfy $p\equiv 2q-m\pmod{N}$.

    We compare the contributions from the following two input Fock components with the same total photon number:
    \begin{subequations}\label{eq:exchanged_input_components}
        \begin{equation}
            \ket{N\ell_1+q}_A\ket{N\ell_2+q}_B,
        \end{equation}
        \begin{equation}
            \ket{N\ell_2+q}_A\ket{N\ell_1+q}_B.
        \end{equation}
    \end{subequations}
    Since a phase rotation of $\pi/N$ is applied to mode $B$, the two input components acquire phase factors $(-1)^{\ell_2}$ and $(-1)^{\ell_1}$, respectively, according to \eqref{eq:phase_rotation_N_component_Fock}. From \eqref{eq:HBS_coherent_state_transformation}, exchanging the two input states in modes $A$ and $B$ leaves the amplitude in the retained mode $C$ unchanged but reverses the sign of the amplitude in the measured mode $D$. This sign change multiplies the $m$-photon component in mode $D$ by a phase factor $(-1)^{m}$. From these results, the sum of the sign factors associated with the two exchanged input components is given by
    \begin{equation}
        (-1)^{\ell_2}+(-1)^{m}(-1)^{\ell_1}=(-1)^{\ell_2}\left[1+(-1)^{L+m}\right],
        \label{eq:exchange_pair_coefficient}
    \end{equation}
    where $L=\ell_1+\ell_2$.

    For $\ell_1\neq\ell_2$, \eqref{eq:exchange_pair_coefficient} shows that the coefficient vanishes when $L+m$ is odd. Thus, $L+m$ must be even for a nonzero contribution. For $\ell_1=\ell_2$, we have $L=2\ell_1$, so $L$ is always even. When the two inputs to the HBS have the same photon number, the odd-photon-number components in output mode $D$ vanish. Therefore, only even values of $m$ have nonzero probability amplitudes. Therefore, we also obtain
    \begin{equation}
        L\equiv m\pmod{2}.
        \label{eq:L_m_parity_equal_case}
    \end{equation}
    Photon-number conservation alone does not determine the parity of $L$. However, due to interference at the HBS, only components for which $L+m$ is even can contribute to a nonzero conditional output. Thus, $L$ and $m$ have the same parity.

    From \eqref{eq:L_m_parity_equal_case}, $L-m$ is even, so we can write $L=m+2u$, where $u$ is an integer. Substituting this into \eqref{eq:remaining_photon_number_after_detection}, we obtain $p=2q+(N-1)m+2Nu$. Therefore, we obtain
    \begin{equation}
        p\equiv 2q+(N-1)m\pmod{2N}.
        \label{eq:conditional_mod_2N_rule}
    \end{equation}
    Thus, the conditional output contains only Fock components with photon numbers equal to $2q+(N-1)m$ modulo $2N$. This selection rule holds for all integers $N\geq2$, regardless of the input amplitude $\alpha$. However, it only specifies which Fock components can appear in the conditional output and does not guarantee that all Fock components satisfying the condition actually appear.

    To examine the relation between the conditional output and a $2N$-component cat state, we define the corresponding ideal cat state. We define $s$ as the remainder when the output photon number is divided by $2N$, with $s=0,1,\ldots,2N-1$. For a measurement outcome $m$, \eqref{eq:conditional_mod_2N_rule} determines $s$ through
    \begin{equation}
        s\equiv2q+(N-1)m\pmod{2N}.
        \label{eq:conditional_2N_cat_phase_index}
    \end{equation}
    For $s$ determined by \eqref{eq:conditional_2N_cat_phase_index}, the ideal $2N$-component cat state with complex amplitude $\beta\neq0$ can be expressed as a superposition of $2N$ coherent states:
    \begin{equation}
        \ket{C_s^{(2N)}(\beta)}\propto\sum_{j=0}^{2N-1}e^{-isj\pi/N}\ket{\beta e^{ij\pi/N}}.
        \label{eq:general_2N_component_cat}
    \end{equation}
    In \eqref{eq:general_2N_component_cat}, the $2N$ coherent states are arranged on the same circle with angular spacing $\pi/N$, and the phase factors of the components are determined by $s$. This state can also be expressed in the Fock basis as
    \begin{equation}
        \ket{C_s^{(2N)}(\beta)}\propto\sum_{\ell=0}^{\infty}\frac{\beta^{2N\ell+s}}{\sqrt{(2N\ell+s)!}}\ket{2N\ell+s}.
        \label{eq:general_2N_component_cat_Fock}
    \end{equation}
    Comparing \eqref{eq:conditional_mod_2N_rule} and \eqref{eq:general_2N_component_cat_Fock}, we find that the conditional output and the ideal $2N$-component cat state contain only Fock components with photon numbers equal to $s$ modulo $2N$. However, this does not imply that their Fock coefficients are equal.

    To examine whether the output is an ideal cat state, we analyze the arrangement and amplitudes of the coherent-state components in the retained mode $C$. We denote the indices of the coherent-state components in modes $A$ and $B$ by $j$ and $k$, respectively. The input cat states defined in \eqref{eq:N_component_cat_A_Fock_expansion} can be expressed as a superposition of coherent states:
    \begin{equation}
        \ket{C_q^{(N)}(\alpha)}\propto \sum_{j=0}^{N-1}e^{-i2\pi qj/N}\ket{\alpha e^{i2\pi j/N}}.
        \label{eq:N_component_cat_coherent_representation}
    \end{equation}
    For $q=0$, all coherent-state components have equal coefficients. For $q\ne0$, each component has a phase factor $e^{-i2\pi qj/N}$. The input state in mode $B$ is obtained by rotating the state in \eqref{eq:N_component_cat_coherent_representation} by $\pi/N$.

    We input the state in \eqref{eq:N_component_cat_coherent_representation} and the same state rotated by $\pi/N$ into the HBS. We denote the coherent-state amplitudes in output modes $C$ and $D$ by $\alpha_C^{(j,r)}$ and $\alpha_D^{(j,r)}$, respectively. Using \eqref{eq:HBS_coherent_state_transformation}, we obtain
    \begin{subequations}\label{eq:HBS_output_amplitudes}
        \begin{equation}
            \alpha_C^{(j,r)}=\frac{\alpha e^{i2\pi j/N}}{\sqrt{2}}\left[1+e^{i(2r+1)\pi/N}\right],
            \label{eq:HBS_output_amplitude_C}
        \end{equation}
        \begin{equation}
            \alpha_D^{(j,r)}=\frac{\alpha e^{i2\pi j/N}}{\sqrt{2}}\left[1-e^{i(2r+1)\pi/N}\right].
            \label{eq:HBS_output_amplitude_D}
        \end{equation}
    \end{subequations}
    Here, $r$ represents the difference between the coherent-state component indices $j$ and $k$, defined as
    \begin{equation}
        r\equiv k-j\pmod{N}.
        \label{eq:r_definition}
    \end{equation}
    With $\theta_r=(2r+1)\pi/(2N)$, the two amplitudes can be written as
    \begin{subequations}\label{eq:HBS_output_amplitudes_trigonometric}
        \begin{equation}
            \alpha_C^{(j,r)}=\sqrt{2}\alpha\cos\theta_r\,e^{i(2\pi j/N+\theta_r)},
            \label{eq:HBS_output_amplitude_C_trigonometric}
        \end{equation}
        \begin{equation}
            \alpha_D^{(j,r)}=-i\sqrt{2}\alpha\sin\theta_r\,e^{i(2\pi j/N+\theta_r)}.
            \label{eq:HBS_output_amplitude_D_trigonometric}
        \end{equation}
    \end{subequations}
    Thus, for a fixed $r$, the coherent-state components in mode $C$ are arranged on a circle with angular spacing $2\pi/N$. The radius of this circle is given by
    \begin{equation}
        \left|\alpha_C^{(j,r)}\right|=\sqrt{2}|\alpha|\left|\cos\frac{(2r+1)\pi}{2N}\right|.
        \label{eq:HBS_output_radius_C}
    \end{equation}
    From \eqref{eq:HBS_output_radius_C}, the coherent-state components corresponding to $r$ and $N-1-r$ are arranged on the same circle. For each value of $r$, varying $j$ gives $N$ coherent-state components with angular spacing $2\pi/N$. The two sets are shifted by $\pi/N$ relative to each other, resulting in $2N$ coherent-state components arranged with angular spacing $\pi/N$.

    When $N$ is even, the coherent-state components are arranged on $N/2$ circles with different radii. When $N$ is odd, the coherent-state components are arranged on $(N-1)/2$ circles with different radii. In addition, a vacuum component appears for $r=(N-1)/2$, since $\alpha_C^{(j,r)}=0$. These results describe the arrangement of the coherent-state components at the HBS output. To obtain the conditional output state after the PNRD measurement, we also need to consider the coefficient of each coherent-state component.

    To determine the coefficients of the generated state, we use the projection of a coherent state in the measured mode onto the $m$-photon state. For an arbitrary complex amplitude $\delta$, we have
    \begin{equation}
        \braket{m}{\delta}=e^{-|\delta|^2/2}\frac{\delta^m}{\sqrt{m!}}.
        \label{eq:coherent_state_photon_projection_general}
    \end{equation}
    Including the input phase factors from \eqref{eq:N_component_cat_coherent_representation}, the conditional output state corresponding to the measurement outcome $m$ is given by
    \begin{equation}
        \ket{\widetilde\psi_m}\propto\sum_{j,r=0}^{N-1}e^{-\frac{|\alpha_D^{(j,r)}|^2}{2}-\frac{i2\pi q(2j+r)}{N}}\left(\alpha_D^{(j,r)}\right)^m\ket{\alpha_C^{(j,r)}}.
        \label{eq:conditional_coherent_state_sum}
    \end{equation}

    \begin{figure}[t]
    \centering
    \includegraphics[width=\linewidth]{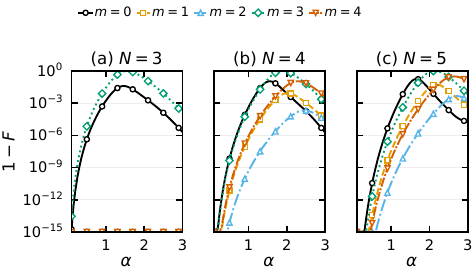}
    \caption{Infidelity $1-F$ as a function of the input amplitude $\alpha$ for (a) $N=3$, (b) $N=4$, and (c) $N=5$. Here, $F$ is the fidelity between the conditional output and the ideal $2N$-component cat state corresponding to the outermost circle of the HBS output. We set $q=0$, and each curve corresponds to a measurement outcome $m=0,1,2,3,4$. For $N=3$, the fidelity is $F=1$ for $m=1,2,4$.}
    \label{FIG02_NS}
    \end{figure}
    
\section{Numerical simulation}
    As shown in the previous section, the photon number of the conditional output is fixed modulo $2N$. In this section, we numerically evaluate how close the conditional output is to an ideal $2N$-component cat state.

    For comparison with the conditional output state, we consider an ideal $2N$-component cat state corresponding to the outermost circle of the HBS output. From \eqref{eq:HBS_output_radius_C}, the outermost circle corresponds to $r=0$ and $r=N-1$. We define the radius of this circle as $\beta_N(\alpha)$:
    \begin{equation}
        \beta_N(\alpha)=\sqrt{2}\alpha\cos\frac{\pi}{2N}.
        \label{eq:outermost_2N_cat_amplitude}
    \end{equation}
    The coherent-state components corresponding to $r=0$ and $r=N-1$ have the same angular arrangement as those in \eqref{eq:general_2N_component_cat}, rotated by $\pi/(2N)$. We numerically evaluate the fidelity between the conditional output state and the ideal $2N$-component cat state. We set $q=0$ and consider $N=3,4,5$ and measurement outcomes $m=0,1,2,3,4$. Fig.~\ref{FIG02_NS} shows the infidelity $1-F$ as a function of the input amplitude $\alpha$.

    For $N=3$, the fidelity is $F=1$ for $m=1,2,4$, and the conditional output becomes an ideal six-component cat state. In this case, the six coherent-state components corresponding to $r=0$ and $r=2$ are arranged on the same circle with radius
    \begin{equation}
        \beta_3(\alpha)=\sqrt{\frac{3}{2}}\alpha,
        \label{eq:N3_six_cat_amplitude}
    \end{equation}
    as obtained from \eqref{eq:outermost_2N_cat_amplitude}. However, for $r=1$, \eqref{eq:HBS_output_amplitude_C_trigonometric} gives $\alpha_C^{(j,1)}=0$, so a vacuum component appears. For $m=1,2,4$, this vacuum component is canceled, and the conditional output is equal to a six-component cat state with the radius given by \eqref{eq:N3_six_cat_amplitude}. In contrast, for $m=0,3$, the vacuum component remains. The conditional output is therefore generally not equal to a six-component cat state.

    For $N=4,5$, \eqref{eq:HBS_output_radius_C} shows that coherent-state components with radii different from that of the outermost circle also appear in the retained mode $C$. Therefore, the conditional output contains contributions from coherent-state components with different radii and is generally not equal to a $2N$-component cat state. The fidelity thus depends on the measurement outcome $m$ and the input amplitude $\alpha$.

    Except for $N=3$ with $m=1,2,4$, where $F=1$, the fidelity is close to 1 for small input amplitudes $\alpha$. As $\alpha$ increases, the fidelity first decreases and then increases again. For small amplitudes $\alpha$, $2N$-component cat states in the same residue class approach the same Fock state with the lowest photon number, even if their amplitudes differ. Therefore, the conditional output is close to the ideal cat state. As the input amplitude $\alpha$ increases, higher-photon-number Fock components cannot be ignored. The resulting differences in Fock coefficients between the conditional output and the ideal cat state reduce the fidelity. The behavior as $\alpha$ increases further can be explained by \eqref{eq:HBS_output_amplitude_D_trigonometric} and \eqref{eq:conditional_coherent_state_sum}. For a fixed $r$, the magnitude of the coefficient multiplying $\ket{\alpha_C^{(j,r)}}$ in \eqref{eq:conditional_coherent_state_sum} is proportional to
    \begin{equation}
        e^{-\alpha^2\sin^2\theta_r}
        \left(\sqrt{2}\alpha|\sin\theta_r|\right)^m.
        \label{eq:conditional_coherent_component_weight}
    \end{equation}
    For $r=0$ and $r=N-1$, which correspond to the outermost circle, $|\sin\theta_r|$ takes its minimum value. Therefore, the exponential factor in \eqref{eq:conditional_coherent_component_weight} decreases more slowly as $\alpha$ increases than for other values of $r$. Thus, for a fixed measurement outcome $m$, the contributions from the other components become smaller relative to those from the outermost circle as $\alpha$ increases. As a result, the components corresponding to the outermost circle dominate for large $\alpha$, and the conditional output approaches an ideal $2N$-component cat state. The fidelity therefore increases again at large $\alpha$ after decreasing at intermediate amplitudes.
    
\section{Discussion}   
    In this paper, we show that by interfering two $N$-component cat states on an HBS and measuring one output with a PNRD, the Fock components of the conditional output are restricted to a single residue class modulo $2N$. For $N=3$, a six-component cat state can also be generated exactly by postselecting the generated state. On the other hand, for arbitrary $N$, the conditional output contains coherent-state components with different radii and therefore does not necessarily coincide with a $2N$-component cat state at finite amplitude. However, for large input amplitudes, the contributions from coherent-state components with radii other than the outermost one become relatively small. Therefore, the conditional output approaches a $2N$-component cat state determined by the outermost circle. These results indicate that, for sufficiently large input amplitudes, the proposed method can approximately generate $2N$-component cat states for arbitrary $N$. 

\begin{backmatter}
\bmsection{Funding} 
    This work was supported by the Advancement of Next Generation Research Projects, Keio University; JST Moonshot R\&D Grant No. JPMJMS256E; JST ASPIRE Grant No. JPMJAP2427; JST SPRING Grant No. JPMJSP2123; and JST COI-NEXT Grant No. JPMJPF2221.

\bmsection{Disclosures} 
    The authors declare no conflicts of interest.
    
\smallskip

%\noindent Here are examples of disclosures:

\bmsection{Data Availability Statement} 
    Data underlying the results presented in this paper are not publicly available at this time but may be obtained from the authors upon reasonable request.
\end{backmatter}

% Bibliography
\bibliography{sample}

\end{document}